\documentstyle[aps,preprint,prb]{revtex}

\begin{document}
\title{1.26 $\mu$m intersubband transitions in In$_{0.3}$Ga$_{0.7}$As/AlAs quantum wells}
\author{C\'{e}sar Pascual Garcia$^{a)}$, Andrea De Nardis, Vittorio Pellegrini$^{b)}$, 
Jean Marc Jancu, and Fabio Beltram} 
\address{Scuola Normale Superiore and INFM, Piazza dei Cavalieri 7, I-56126 Pisa, Italy}
\author{Bernhard H. M\"ueller$^{c}$, Lucia Sorba$^{d}$, and Alfonso Franciosi$^{e}$}
\address{Laboratorio Nazionale TASC-INFM, I-34012 Trieste, Italy}
\footnotetext[1]{Current address: Departamento de F\'{\i}sica de materiales, Universidad
Aut\'{o}noma de Madrid, E-28049 Madrid, Spain}
\footnotetext[2]{E-mail: vp@sns.it}
\footnotetext[3]{Present address: Institut f\"ur Halbleitertechnologie, Universit\"at Hannover, 
D-30167, Hannover, Germany}
\footnotetext[4]{Also with Dipartimento di Fisica, Universit\`a di Modena, I-41100 Modena, Italy}
\footnotetext[5]{Also with Dipartimento di Fisica, Universit\`a di Trieste, I-34127 Trieste, Italy}
\maketitle
\begin{abstract}
We observed room-temperature intersubband transitions at 1.26 microns 
in n-doped type-II In$_{0.3}$Ga$_{0.7}$As/AlAs strained quantum wells. 
An improved tight-binding model was used to optimize the structure parameters in order
to obtain the shortest wavelength intersubband transition ever achieved in a semiconductor system.
The corresponding transitions occur between 
the first confined electronic levels of the well following mid-infrared 
optical pumping of electrons from the barrier X-valley into the well ground state. 
\end{abstract}

\newpage

Intersubband optical transitions (ISTs) between quantized electronic
levels in semiconductor quantum wells (QWs) are at the basis of
the realization of novel mid-infrared semiconductor detectors~\cite{Levine} and lasers.\cite{Faist} 
In recent years much attention was given to the implementation of systems exhibiting ISTs with shorter 
wavelengths extending to the near infrared. 
In fact, thanks to the ultrafast carrier-relaxation processes 
characteristic of intersubband scattering such high-energy ISTs can
find many important applications for high-speed device operation
at short wavelengths~\cite{Elsaesser,West} and for ultrafast 
all-optical modulation schemes~\cite{Noda} that would require 
similar wavelengths for both 
interband and intersubband transitions. The achievement of short-wavelength intersubband transitions 
is basically connected to the choice of suitable material combinations yielding 
adequately large conduction-band offsets. In particular, systems composed by 
narrow In$_{x}$Ga$_{1-x}$As/AlAs strained QWs 
were extensively studied  ~\cite{Marinet,Hirayama} 
and yielded wavelengths as short as 1.59$\mu$m~\cite{Sung} 
and 1.55$\mu$m~\cite{Smet}, thanks to the large
In$_{x}$Ga$_{1-x}$As/AlAs conduction-band offset and large
tunability of the IST energies. Picosecond intersubband
relaxation in this material system for the case of ISTs at around 1.8$\mu$m 
was also recently demonstrated.\cite{ghislotti} 
The use of InGaAs/AlAsSb compounds has shortened these wavelengths down to
1.45$\mu$m~\cite{Mozume} which represents the highest-energy IST reported so far.
Additionally, GaN/AlGaN narrow QWs have been studied very recently and 
displayed intersubband transition at 1.77$\mu$m\cite{gmachl}. This latter material combination may offer important avenues for the achievement of very short-wavelength ISTs.
\par
In our recent work we demonstrated that n-doped In$_{x}$Ga$_{1-x}$As/AlAs heterostructures
allow a large tunability of IST energies and can yield intersubband wavelengths well
below 1.5$\mu$m.\cite{jmj1} We also showed, however, that  
quantum-size effects associated with well thickness
and indium mole fraction values can induce a type
I$\rightarrow$type II transition as the QW width is decreased. 
This happens when the quantum-confined $\Gamma$ conduction-band minimum 
localized in the In$_{x}$Ga$_{1-x}$As well is pushed above the bulk AlAs X 
conduction-band minimum, leading to an indirect band-gap configuration. 
In this case, at equilibrium, electrons populate the X valley in the barrier. 
Intersubband absorption thus occurs in the mid-infrared and is associated with indirect 
electronic transitions from the X (barrier) to the $\Gamma$ (well) states. 
Observation of short-wavelength direct 
intersubband transitions below 1.5\,$\mu$m 
is hindered by the lack of electrons in the lowest $\Gamma$-like subband of the QW. 
\par
In this letter we shall focus on n-doped type-II In$_{x}$Ga$_{1-x}$As/AlAs multiple QW 
structures with narrow well widths (few monolayers) and we shall demonstrate that it is 
possible to photo-activate
the very short-wavelength direct intersubband absorption associated to confined $\Gamma$ states in the QWs 
by photo-pumping electrons out of the barrier X levels and into 
the lowest $\Gamma$-like state of the wells. The approach used here  
allowed us to circumvent limitations associated 
with the type-II character of these ultrathin QWs. As a result, we obtained 
room-temperature IST at 1.26 microns (0.98 eV) 
in a 5-monolayer(ML)-thick In$_{0.3}$Ga$_{0.7}$As/AlAs multiple-QW structure
(1 ML $\approx $ 3 $\rm{\AA}$). This observation sets the new record for the 
shortest-wavelength IST achieved in a semiconductor system.
\par
In order to identify the optimal In$_{x}$Ga$_{1-x}$As/AlAs configuration 
for short-wavelength ISTs with large dipole moment,
band-structure calculations were performed within the tight-banding (TB)
approximation using a 40-band empirical $sp^3d^5s^*$ nearest-neighbor model
that includes spin-orbit coupling.\cite{jmj0} 
The usefulness of the present TB model was recently demonstrated in the 
calculation of the optical properties of several III-V 
semiconductor quantum wells and superlattices \cite{rs} 
including the InGaAs/AlAs heterostructure system here of interest.\cite{jmj1} 
\par
Figure 1 (left panel) shows the calculated squared dipole matrix element $E_{P}$ (in eV)
of the predominant IST between 
the $\Gamma$-like conduction-band states, named c1 and c2, 
as a function of alloy composition and well thickness
($E_{P}=\frac{2}{m_{o}} \langle c1 \mid P_{z} \mid c2 \rangle^{2}$, 
where $m_{o}$ is the free electron mass).  
The corresponding intersubband transition energies are 
displayed in the right panel of Fig.\,1.
These calculations suggest that useful, very short-wavelength 
intersubband transitions can indeed 
be achieved in this material system with ultrathin QWs. 
In particular, samples with well width of 5\,MLs emerge as 
most promising candidates for short-wavelength intersubband 
absorption. In fact, 4-ML-wide QWs that present even larger 
IST energies do not yield an efficient localization 
of the c2 $\Gamma$-like state in the InGaAs QW region, which reflects in
the low intersubband dipole moments reported in Fig.\,1\,(left panel). 
Setting the QW thickness at 5\,ML, however, demands much care in the 
choice of the alloy composition. At large In content (x $>$ 0.6), 
these structures exceed the critical thickness as calculated 
following Matthews and Blakeslee's mechanical-equilibrium model.\cite{mb} 
From our calculations, the best In concentration value for the realization of 
short-wavelength IST is estimated at around x = 0.3. 
It must be noted that this well width (5ML) and
In concentration lead to a type-II configuration and intersubband 
absorption in the mid-infrared\cite{jmj1} as pointed out above. 
We shall demonstrate, however, that
photoinduced occupation of the confined $\Gamma $-like subband of 
the QW enables the observation of the c1-c2 intersubband absorption.
\par
The sample used in this study was grown by molecular beam epitaxy 
on a GaAs(001) substrate and consists of a 0.5 $\mu$m GaAs buffer layer 
followed by 30 In$_{0.3}$Ga$_{0.7}$As QWs 5-ML wide, grown at 540$^{o}$C with
a 30 s growth interruption at each interface, 
separated by 108-$\rm \AA$-thick (36ML) AlAs barriers. Wells were Si doped to 
1.4$\times$10$^{19}$cm$^{-3}$. The growth was
concluded with a 10-nm-thick GaAs cap layer. Well and barrier thickness were
verified by transmission electron microscopy. 
The sample was fabricated in a multipass waveguide geometry with 
45$^{\circ}$ polished mirror facets and placed in a 
Fourier Transform Infrared spectrometer (FTIR) for the mid-infrared 
absorption measurements.
\par
Figure 2 shows the room-temperature mid-infrared absorption of the sample. 
The spectrum was acquired with z-polarized incident light 
(perpendicular to the QW plane) and normalized 
with respect to the in-plane polarized transmission signal. 
The absorption peak is centered at 98 meV ($\approx $ 12 $\mu$m) 
and corresponds to the X-$\Gamma $ indirect transition.\cite{jmj1} 
This transition is induced by $\Gamma $-X intervalley mixing and 
is negligible for in-plane polarization owing to the reduced 
overlap between the envelope functions of the two states 
(see inset of Fig.\,2). It must be noted that this selection rule is peculiar
to our system; in short-period type II superlattices, for instance, where
the X state is quantum-confined and therefore 
the wavefunction overlap is larger, 
the X-$\Gamma $ intersubband transition is basically polarization insensitive.\cite{feni}

The  energy-level diagram of the heterostructure as derived within our TB approximation 
is shown in the inset of Fig.\,3. The c1-c2 intersubband transition
is calculated at 0.94 eV ($\approx $1.3 $\mu$m). 
\par
In order to observe the short wavelength c1-c2 IST we used 
the radiation from a CO$_{2}$ laser to pump electrons out of the
X level to the c1 $\Gamma $-like ground state in the well. 
During the experiment the sample was illuminated on the same spot (a few millimeters wide) 
with both the CO$_{2}$ laser 
and a halogen lamp for the transmission measurements.
The CO$_{2}$ laser radiation was tuned at 110 meV (within the X-$\Gamma $ 
absorption peak shown in Fig.\,2) and focused on the sample 
surface with an intensity of $\approx $ 20 W/cm$^2$.
This high power density was motivated by the low oscillator strength 
of the indirect intersubband transition (f=0.001 compared with f$\approx $3.3
for the c1-c2 transition).
The transmission signal was then dispersed by a 32-cm-long monochromator 
and detected by a Ge detector and conventional lock-in techniques. 
\par
Figure\,3 shows the photo-activated intersubband absorption at room temperature. 
The peak is centered at 0.98 eV, $\approx$ 1.26 $\mu$m, in good agreement 
with the TB calculation results (see Fig.\,1) and has a full width at half 
maximum of 60\,meV which is largely due to well-width and alloy fluctuations. 
Data plotted were obtained by careful normalization to the 
transmission background signals obtained {\it without} concurrent CO$_{2}$ excitation.
In order to rule out the influence of local heating effects on the observed near-infrared absorption peak 
we checked the impact of a different normalization procedure.
To this end, we acquired transmission data by using as 
normalization background signal the transmitted intensity under 
concurrent laser irradiation 
at 135\,meV with an intensity of $\approx $ 20 W/cm$^2$. 
(The excitation frequency lays out of the absorption peak shown in Fig.\,2.)
The same absorption peak was observed, confirming 
unambiguously that photo-pumping across the mid-infrared X-$\Gamma $ transition is
the dominant mechanism that yields electrons in the well. 
We also found that the c1-c2 absorption peak 
increased with excitation power further highlighting the 
photoassisted nature of the observed signal. 
\par
In conclusion, we have reported the experimental observation of 
room-temperature intersubband transition at 1.26 microns in type-II 
In$_{0.3}$Ga$_{0.7}$As/AlAs quantum wells.
For this experiment, electrons were photo-pumped 
from the X valley located in the AlAs barrier to the lowest 
confined $\Gamma $-like state in the well by 
means of CO$_{2}$ laser irradiation.
Further engineering of InGaAs/AlGaAs heterostructures 
that will take advantage of quantum confinement of X-valley levels in 
the barrier will open the way to the exploitation of very short wavelength 
IST in optoelectronic devices even without 
the need for MIR-photoexcitation. 
\par
This work was supported in part by the Consiglio Nazionale delle Ricerche (CNR) 
within the CNR-Scuola Normale Superiore framework agreement and by MURST. Two of the 
authors (CPG and BHM) acknowledge financial support from the European Commission 
under the ERASMUS and TMR programs, respectively. One of the authors (ADN) acknowledges 
IFAM-CNR for financial support. We thank A. Parisini and P.G. Merli for the transmission 
electron microscopy analysis.

\begin{figure}
\caption{Lowest dominant intersubband transitions between $\Gamma $-like 
conduction states (c1,c2) as a function of well thickness (in monolayers, ML) and 
alloy composition. Left Panel: Calculated dipole matrix element squared ($E_{P}$ in eV). Right Panel: 
Calculated intersubband transition energy 
($\Delta E$). The oscillator strength (f) for each transition is given by f=$E_{P}$/$\Delta E$.}
\end{figure}

\begin{figure}
\caption{Room-temperature mid-infrared intersubband absorption of 5-monolayer 
In$_{0.3}$Ga$_{0.7}$As quantum wells confined by AlAs barriers. The inset shows 
the polarization dependence of the intersubband absorption peak. Zero degrees 
corresponds to incident light polarized along the growth (z) direction.}
\end{figure}

\begin{figure}
\caption{Room-temperature near-infrared intersubband absorption of 5-monolayer 
In$_{0.3}$Ga$_{0.7}$As quantum wells confined by AlAs barriers during CO$_{2}$ laser irradiation at 110 meV and $\approx $20 W/cm$^{2}$ incident intensity. Each data point corresponds to 30-s integration time.
Transmission signals were normalized to signals obtained without CO$_2$ laser irradiation. 
The inset displays the calculated band alignment of the heterostructure. The calculated c1-c2 
intersubband transition energy is 0.94\,eV. The dashed line indicates the CO$_{2}$-induced 
electron photopumping across the X-$\Gamma $ transition.}
\end{figure}

\end{document}